# Bhāsvatī. of Śatānanda: In the Pages of Mystery

Sudhira Panda
Plot-297, Nayapalli,
Behera Sahi, Bhubaneshwar,Odisha.

## Abstract:

Śatānandācārya, the astronomer and mathematician of 11th century, was born in 1068 C.E. at Purusottamdham Puri (Jagannath Puri) of Odisha wrote the scripture *Bhāsvatī* in 1099 C.E.[1]. This scripture has significant contribution to world of astronomy and mathematics. Śatānanda has adopted Centesimal System for the calculation of position and motion of heavenly bodies, which is similar to the present day Decimal System[2]. The treaties got the recognition of a scripture of Karana grantha[3]. Commentary of this work has been made by different persons during different times of history[4]. Though it is found to be remade in almost once in each century and was well known all over India and abroad. Presently it is completely lost and no reference is available in ongoing works. The main aim of this paper is to outline and bring to the notice of a wider audience- the genius of Śatānanda and his contribution to the world of astronomy and mathematics.

## Introduction:

The history of development of mathematics in India is as old as Vedas. From the prehistoric days mathematics began with rudiments of metrology and computation, of which some fragmentary evidence has survived till date. The sacred literature of the Vedic Hindus- the *Saṁhītās*, the *Kalpas* and the *Vedāṅgas* contain enough materials, which prove the mathematical ability of those pioneers who developed this class of literature. Those pioneers, mostly astronomers, were using mathematics as an instrument for calculation of position of stars and planets. Rather one can say that such calculation of heavenly bodies, their positions and movements (Astronomy) urged for the origin of mathematics i.e. addition, subtraction, multiplication and division, so also fractions. The division of the days, the months, the seasons contemplated an idea of fractions.

In all ancient calculations the astronomers assigned 360 *aṁśa* for one cycle, since 360 is the smallest number divisible by the integers 1 to 10 excluding 7. The trend is still implemented in the present day calculations. However in late 11th century an astronomer Śatānanda born in Odisha tried to make a deviation from the ongoing mathematical research and was successful in his attempt. He had converted all cyclic calculations into multiples of hundred for convenience. He had used 1200 *aṁśa* while calculating the positions and motions of planets with respect to 12 constellations and used 2700 *aṁśa* while calculating the positions and motions of the Sun and Moon with respect to 27 Nakṣatras.

The scripture *Bhāsvatī* of Śatānanda has introduced very simple methods to calculate celestial parameters without using trigonometric functions. Therefore it was appreciated by the society and it spread all over north India, though many astronomers like Samanta Chandra Sekhar considered it as an approximate calculation. Transformation of *aṁśa* (degrees) into *śatāṁśa* (multiple of hundred) was the greatest achievement of Śatānanda of 11$^{th}$ century recorded in *Bhāsvatī.* There is a claim exists that this mathematical calculation was the initial form of the modern day decimal system calculation[2].The commentary of this work was made almost in each century in history of India and abroad. In the present day research, this reference is completely ignored by the mathematicians and astronomers of our country. This pioneering piece of work of Śatānanda has been very little known even in the learned society of his native place Odisha.

In this paper the mathematical calculations where Śatānanda had introduced (i) Centesimal fractions and (ii) converted the *aṁśa* (degrees) into *śatāṁśa* (multiple of hundred) have been explained. Section I deals in introduction about the history of mathematical Science before Śatānanda wrote *Bhāsvatī*, Section II, deals in the historical details regarding Śatānanda and Comments and Commentaries on *Bhāsvatī*, Section III, explained the mathematics, Śatānanda had introduced in his text *Bhāsvatī* and the explanations to it. Section-IV the conclusion and the future plan.

**Section I: Śatānanda**: From the history of Odisha it is known that Śatānanda might have been a courtier in the period of Keśāri Dynasty (474 C.E.-1132 C.E.). In that period, many constructive works were done. The kingdom was peaceful and patronage was given to scientists and architects. Establishment of Cuttack city, the then state capital had been made in that period. Besides this the stone embankment on the river *Kāṭhajoḍi* and *Aṭharanalā* bridge of Śrikhetra Purī were the significant achievements of that time. *Bhāsvatī* of Śatānanda was the greatest achievement of Keśāri dynasty.

Śatānanda wrote this scripture, which was a guideline to make *Pañcāṅga* (calendar) for the benefit of performing rituals in Jagannātha temple, Puri. Since *Pañcāṅga* (calendar) has an important role in Hindu society. Śatānanda made the calculation of heavenly events of heavenly bodies accurately. Hence there was a saying in Varanasi (the then knowledge center of India) -- **ग्रहणे भास्वती धन्या** (*Bhāsvatī* is the best book to predetermine eclipses.). It is also enlightening to know that the great Hindi poet Mallik Muhammad Jayagi praised *Bhāsvatī* in his book as[1]

भास्वती औव्याकरनपिङ्गलपाठपुराण।
वेदमेद सो वात कहि जनुलागेहिय वान ||

This shows the popularity of Bhāsvatī in the society.

---

## Section II: Commentaries:

There is a commentary on *Bhāsvatī* written in Śaka 1417 by Anirudddha of Varanasi from which it appears that there were many other commentaries on it written before[1].

Mādhava a resident of Kanauja (Kānyakubja) wrote the commentary of *Bhāsvatī* in *Śaka* 1442. Another commentary of this scripture was written in Śaka 1607 by Gaṅgādhara. The author of the commentary written in *Śaka* 1577 is not known. According to the Colebrooke, the commentary of Balabhadra born in Jumula region of Nepal was written in *Śaka* 1330[2]. From the catalogue of Sanskrit books prepared by Aufrecht, the title of this commentary appears to be *Bālabodhinī.* This book was the first mathematics text book in Nepal[6], since the mathematical operations like Additions, Subtractions, Multiplications and Divisions are explained explicitly in *Bhāsvatī*. According to Aufrecht's Catalogue there are following additional commentaries on *Bhāsvatī karaṇa*: *Bhāsvatī karaṇapaddhati*; *Tatvaprakāśikā* by Rāmakṛuṣṇa, *Bhāsvatīcakraraśmyudaharana* by Rāmakrṛṣṇa, *Udāharaṇa* by Śatānanda, *Udāharaṇa* by Vṛundāvana. Similarly, there are commentaries by Achutabhaṭṭa, Gopāla, Cakravipradāsa, Rāmeśvara, Sadānanda and a "*Prakrit*" commentary by Vanamāli. Very recently it is found that there was commentary of this scripture with examples in Odia by Devīdāsa in *Śaka* 1372 from the State Museum, Odisha. This is a well explained book on mathematics and heavenly phenomena calculated in *Bhāsvatī*.

Most of these commentators hail from Northern India. The author of History of Indian Astronomy Sankar Balakrishna Dixit regrets that this great work is not known and there is no reference of this work has been uttered in any research presently.

The copy of these commentaries are presently available in the library of (i) Alwar (Rajasthan), (ii)Asiatic Society, Bengal(Kolkata), (iii)India Office Library (London), (iv)Rajasthan Oriental Research Institue(Jodhpur), (v)Saraswati bhavan Library(Banaras), (vi)Visveswarananda Institute(Hosiarpur), (vii)Bhandarkar Oriental Research Institute(Pune)[5].

## Section III: Contents of *Bhāsvatī* :

*Bhāsvatī* contains 128 verses in eight *Adhikāras* (chapters). Those are (i) *Tīthyādi dhruvādhikāra*(Tithi *Dhruva*), (ii) *Grāhadhruvādhikāra* (*Graha Dhruva*), (iii) *Pañcāṅga spaṣṭādhikāra*(Calculation of Calendar), (iv) *Graha spaṣṭādhikāra* (True place of Planets), (v) *Tripraśnādhikāra*(Three problems: Time, Place and Direction),(vi) *Chandragrahaṇādhikāra* (Lunar Eclipse), (vii) *Sūryagrahaṇādhikāra* (Solar Eclipse),(viii) *Parilekhādhikāra* (Sketch or graphical presentations of eclipses)[1].

The first sloka of his scripture Śatānanda acknowledged the observational work of Varāhamihira which he has used in his calculation. He also claims that his calculations are as accurate as *Sūrya Siddhānta* though the methods of calculation are completely different. The *Śloka* is as follows:

**अथ प्रवक्ष्ये मिहिरोपदेशाच्छ्रीसूर्य्यसिद्धान्तसमं समासात् ।**

Indian astronomers have differed on the rate of precession during different periods with respect to the 'zero year'. The accumulated amount of precession starting from 'zero year' is called ***ayanaṁśa***.

There are different methods to calculate the exact amount of *ayanāṁśa*. (i)The *Siddhāntas* furnish rate for computing it, which is in principle the same as the method of finding the longitude of a star at any given date by applying the amount of precession to its longitude, at some other day. (ii) Defining the initial point with the help of other data such as the recorded longitudes of the stars, its present longitudes from the equinoxial point may be ascertained. (iii) Knowing the exact year when the initial point was fixed, its present longitude *ayanāṁśa* may be calculated from the known rate of precession. However it is so happen that the result obtained by these three methods do not agree. Śatānanda has his own method of calculation which is very simple but considered to be approximate.

*Bhāsvatī*, has assumed *Śaka* 450 (528 C.E.) as the Zero precession year and 1 minute as the rate of precession per year. However Jogesh Chandra Roy in his 61 page introduction to *Siddhānta Darpaṇa* claims that the zero precession years adopted in Bhasvati is Saka 427(505 C.E.). He got this number by making the reverse calculation. The calculation of *ayanāṁśa* has been explained in first sloka of 5th chapter *Tripraśnādhikāra*. The meaning of this *śloka* is :

Subtract 450 from the past years of *Śālivāhana* (Śaka) and then divide it with 60.The quotient is the *ayanāṁśa* (precession). Add *ayanāṁśa* with *ahargaṇa* to bring the proof of day night duration.

Example: If we will subtract 450 from Śaka 1374, it will be 924. Dividing 924 with 60 becomes 15| 24. By adding this value with *ahargaṇa* 27 the result becomes *sayana dinagana* as 42|24.

The table for 'zero *ayanāṁśa*' year and annual rate of precession adopted in different scriptures is given below.

**TABLE-1: ZERO AYANAMSA YEAR AND ANNUAL RATE OF PRECISSION**

| *Siddhānta* | Annual rate of precession | Zero year of equinox in C.E. |
|---|---|---|
| *Sūrya Siddhānta* | 54" | 499 |
| *Soma Siddhānta* | 54" | 499 |
| Laghu-*Vasiṣṭha Siddhānta* | 54" | 499 |
| *Grahalāghava* | 60" | 522 |
| *Bhāsvatī* | 60" | 528 |
| *Bṛhatsaṁhitās*, Manjala(Quoted by Bhāskara-II) | 59.9" | 505 |
| Modern data | 50.27 | |

**TABLE-2: SIDEREAL PERIODS IN MEAN SOLAR DAYS**

| Planets | European Astronomy | *Sūrya Siddhānta* | *Siddhānta Śiromaṇi* | *Siddhānta Darpana* | *Bhāsvatī* |
|---|---|---|---|---|---|
| Sun | 365.25637 | 365.25875+00238 | 365.25843+00206 | 365.25875+00238 | 365.25865+00228 |
| Moon | 27.32166 | 27.32167+00001 | 27.32114-00052 | 27.32167+00001 | 27.32160+00006 |
| Mars | 686.9794 | 686.9975+0181 | 686.9979+0185 | 686.9857+0063 | 686.9692-0102 |
| Mercury | 87.9692 | 87.9585+0107 | 87.9699+0007 | 87.9701+0009 | 87.9672-0020 |
| Jupiter | 4332.5848 | 4332.3206-2642 | 4332.2408-3440 | 4332.6278+0430 | 4332.3066-2782 |
| Venus | 224.7007 | 224.6985-0022 | 224.9679-0028 | 224.7023+0016 | 224.7025+0018 |
| Saturn | 10759.2197 | 10765.7730+6.5533 | 10765.8152+6.5955 | 10759.7605+5408 | 10759.7006+0599 |

It is seen from the above table-2 that the sidereal periods of the Sun and moon calculated in *Bhāsvatī* is almost same as in *Sūrya Siddhānta* and has materially advanced upon it as regards the periods of the other planets[8]. Having regard to the comparatively slow motion of Jupiter and Saturn.

Śatānanda might be very clever to introduce a new calendar from the date he dedicated his work *Bhāsvatī* for the benefit of the Society. Many calendars were introduced by that time. Those were *Śakābda, Gatakali, Hijirābda, Khriṣṭābda* (C.E.) and so on. But Śatānanda took *Śakābda* and Gatakali as his reference calendar and initialized *Śāstrābda*. He explained the method to convert *Śakābda* and *Gatakali* into *Śāstrābda* in the 1st chapter i.e.*tithyādi-dhruvādhikāra*. The śloka and its exact translation are given below.

गतकळिः प्रकारान्तरेण शास्त्राब्दविधिश्च-

**शाको नवाद्रीन्दुकृशानुयुक्तः कलेर्भवत्यब्दगणस्तुवृत्तः।**

**वियन्नभोलोचनवेदहीनः शास्त्राब्दपिण्डः कथितः स एव॥1.२॥**

*Gatakali* can be ascertained by adding 3179 to *Śakābda*. Subtract 4200 from Gatakali, the result is known as ***Śāstrābdapiṇḍa***.)

Example: The above method has been implemented to convert the present year 2017 C.E. to *Śāstrābda*.

The present year 2017 C.E.-78= 1939 *Śakābda*.

*Śakābda* 1939+ 3179=5118 *Gatakali*

*Gatakali* 5118 – 4200= 918 *Śāstrābda*.

Hence as per the record, *Bhāsvatī* has been written in 1099 C.E. and 918 years have been passed.

---

However, in this article I have referred the *ṭīkās* made in Śaka 1374(1452 C.E.) i.e. Śāstrābda 353 . Therefore all the examples mentioned here are in Sastrābda 353.

In this chapter *tithyādi-dhruvādhikāra* Śatānanda had given the method to determine solar days (*tithi*) and longitude (*dhruva*) of nine planets Sun (*Ravi*), Moon (*soma*), Mars (*Maṅgala*), Mercury (*Budha*), Venus (*Śukra*), Jupiter ( *Bṛhaspati*), Saturn (*Śani*), and *Rāhu*, *Ketu* (the shadow planets). He started his calculation from Sun (Ravi).

In the same chapter-1 śloka 4 and 5 he had given an empirical method for determining the longitude (*dhrūvāṅka*) of Sun. The *ślokas* are mentioned below.

*संवत्सरपालक-शुद्धि सूर्य्यध्रुवविधय :-*

***अथ प्रवक्ष्ये मिहिरोपदेशाच्छ्रीसूर्य्यसिद्धान्तसमं समासात् ।***<br>***शास्त्राब्दपिण्डैः स्वरशून्यदिग्घ्नस्तानाग्नियुक्तोऽष्टशतैर्विभक्तः ॥१.४॥***<br>***लब्धन्नगैः शेषितमङ्गयुक्तः सूर्यादिसंवत्सरपालकः स्यात् ।***

***शेषं हरे प्रोज्झ्य पृथग् गजाशा लब्धं  रवेरौदयिको धृवः स्यात्।॥१.५॥***

*Multiply 1007 to Śāstrābda and add 349 and divide by 800 add 6 to the quotient and divide the quotient by 7. The reminder is the saṁvatsarapālaka of Sūrya. By subtracting it from the* *divisor Sudhi comes.*

*Keep this value in two places. Divide by 108 to the digit of one place. That is the dhrūva (longitude) of madhyama Sūrya. Quotient  should be taken up to three places* .)

**Mathematically:**

*Śāstrābda* 919X 1007= 925433 +349 =925782,

925782 ÷ 800 =1157, with reminder 182 (1)

1157 + 6= 1159÷ 7=166, with reminder 1 =>the fourth *graha* (planet) from Sun, i.e. Soma (Moon) is the *Saṁvatsara pālaka*

From (1) 800 – reminder 182 = 618 *Śuddhi*

*Śuddhi* 618÷ 108 = 5 *aṁśa* , with reminder 78

78X60 = 4680  ÷ 108 = 43 *kalā*, with reminder  36

36 X 60 = 2160 ÷ 108 = 20 *vikalā*

So the *dhrūvāṅka* ( longitude) of morning Sun on *Caitra Śukla Pūrṇimā* (Full moon day of the month of *Caitra*) is 5|43|20 *aṁśa* or 5 *aṁśa* 43 *kāla* 20 *vikāla*.  In *Bhāsvatī,* Śatānanda first initialized the position of planets on *Caitra Śukla Pūrṇimā* and then calculated the rate of motion, position and time taken by the planets to complete one rotation in its orbit from the ahargana (the day count ), unlike other *siddhānta*s including *Sūryasiddhānta* which has taken the starting point approximately from the date of the beginning of the civilization( i.e. 6 *manu*+7 *Sandhi*+27 *mahāyuga*+3 *yuga*+present years elapsed from *kaliyuga*) for this purpose. Therefore the number is huge so there is every possibility to make mistake. With all these simplifications *Bhāsvatī* still regarded has an authority for the calculation of eclipse.

**Implementation of *Śatāṁśa***:

Ancient Indian astronomy believes the effect of 12 constellations and 27 *Nakṣatras* on the human life. They took 360 *aṁśa* approximately for one rotation, in 365 days, approximately $1^0$ for one day and specified 30 *aṁśa*  to each constellation and 40/3 *aṁśa* to each star out of 12 constellations and 27 *Nakṣatras* respectively.

Śatānanda very cleverly multiplied 30/4 to 360 *aṁśa* to make it a multiple of hundred without losing the generality.

360X 30/4= 2700 *aṁśa*

Hence each constellation has 225 *aṁśa*  and each *nakshatra* has 100 *aṁśa*.

He adopted 2700 *aṁśa* for the calculation of motion (Sphuṭa gati) of Sun, Moon and Rāhu and Ketu the shadow planets. However he adopted 1200 *aṁśa* for the calculation of motion of other planets like Mars (*Maṅgala*), Mercury (*Budha*), Venus (*Śukra*), Jupiter( *Bṛhaspati*) and Saturn(*Śani*) by taking each constellation as 100 *aṁśa* and 400/9 to each nakshetra to avoid dealing with huge number.

In Chapter –IV (*Graha spaṣṭādhikāra*), Śatānanda introduced *śatāṁśa* while determining the positions of planets.

Example: In *śloka* 4.10 he explained the position of the shadow planets *Rāhu* and *Ketu* as follows.
राहुकेतु स्पष्ट विधि:-

**अहर्गणं वेदहतं दशाप्तं ध्रुवार्द्धयुक्तं भवतीह पातः।**
**खखागनेत्रान्तरितो मुखं स्याच्चक्रार्द्धयुक्तं स्फुट राहुपुच्छः॥४.१०॥**

(Multiply *dinagaṇa with 4 and then divide by 10. Add the quotient with last given dhruva (longitude). Subtract it from 2700. That is Rāhu. Again by dividing the given number by 225 rāśi (constellation) of Rāhu will come.*

*Then by adding cakrārdha 1350 to Rāhu, ketu comes. And by dividing the position number of Ketu by 225, rāśi (constellation) of ketu can be determined. )*

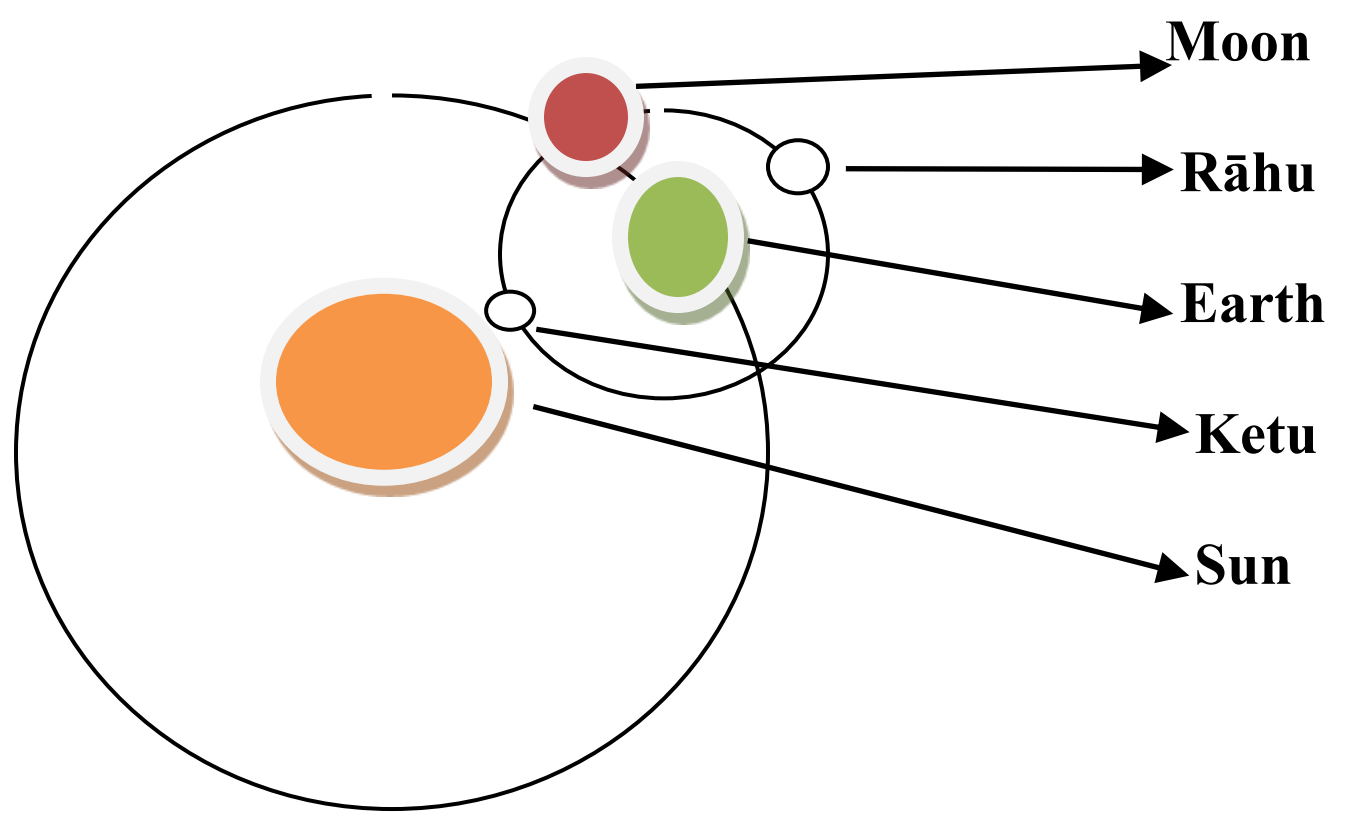

[ Fig-1: The position of Sun, Earth and Moon for the calculation of eclipse time.]

Mathematically: *Ahargaṇa* 27 X 4 =108 ÷ 10 = 10|48 |0

Longitude of *rāhu* 5130|25| 39 ÷ 2= 2565|12|49 + 10|48 |0 = 2576|0|49

2700 - 2576|0|49 = *Rāhu* 123|59 |11

*Rāhu* 123|59 |11 ÷ 225 = 0|16|39|47 is the *rāśi* of *Rāhu*

Again *cakrārdha* 1350 + *Rāhu* 123|59|11= *ketu* 1473|59|11

Here Śatānanda took *cakrārdha* (half rotation) as 1350, as one *cakra* (rotation) is 2700 *aṁśa* .

It was known that Rāhu and Ketu points are opposite to each other($180^0$ apart) in a circle and when Moon is near Rāhu point then there is a chance of getting Lunar eclipse and when is on Ketu point Solar eclipse occurs.

*Ketu* 1476|59|11 ÷ 225 = 6|16|39|47 is the *rāśi* of Ketu

This is an example of implementation of *Śatāṁśa* in *Bhāsvatī*.

Implementation of *Śatāṁśa* had a significant role in predetermining solar and lunar eclipses. It is because (i) 2700 *aṁśa* is a very big number in comparison to 360 *aṁśa* ,(ii) assigning 100 *aṁśa* to each *nakṣatra* or constellation could avoid many error while taking fractions.

## Section-IV

**In this section we want to show the simplified method to calculate time from gnomonic shadow introdeced in** *Bhāsvatī*. **Calculation of time from the Gnomonic (*Śaṅku*) Shadow explained** in *Bhāsvatī*:

Example : Calculation of time on 15th June of this year, when the shadow of the 12 unit gnomon becomes 15 units.

Ans: Here the equinoxial day is 23rd March.

So number of days elapsed= 8 days of March+ 30 days of April +31 days of May +15 days of June= 84 days

or 30 days Aries +30 days Tarus + 24 days Gemini=84 days

Now to calculate ***carārdha litā***

for the month of Aries= 30+30/2 = 45

for the month of Taurus= 30+30/6 = 35

for the month of Gemini= 24/2 = 12

So *carārdha litā* = 45 + 35+ 12 = 92 =Danda 1|32 lita on the day required

***Dinārdha***= 15 +1|32= 16|32 *daṇḍa*

**To calculate *Madhya Prabhā***

*Carārdha litā* 92 X6= 552/10 = 55|12

552- 55|12 =( 496|48 )/10 =49|41

On 15th June Sun is in northern hemisphere. So the above number should be kept as it is.

Now 49|41 – *Akṣa* 44|43 =4|58 ---› *Madhya prabhā*

Here the gnomonic shadow or *Iṣṭa chāyā* =15|0 *aṅgula* X 10 =150 +100 = 250

250 – *Madhya prabhā* 4|58= 245|02 = 245 X 60 +2 = 14702 ---› *Śaṅku*

now *Dinārdha* 16|32 = 16 X60 +32 =992

992 X 100 =99200

99200/14702 = *daṇḍa* 6|45 *litā*

Now we have to convert it modern time.

*daṇḍa* 6|45 *litā* ~ 2 hours and 42 minutes

As we know in Indian astronomy day starts from the sunrise.

Dinardha on 15th June is 16|32 ~ 6 hours and 22 minutes= $6^h\ 22^m$

Midday at $87^0$ longitude = $12^h - 14^m = 11^h\ 46^m$

$11^h\ 46^m - 6^h\ 22^m = 5^h\ 24^m$ ---› time of Sunrise

$5^h\ 24^m + 2^h\ 42^m = 8^h\ 06^m$ ---› is the required time when the shadow of 12 *aṅgula Śaṅku* becomes 15 *aṅgula*.

**Physical explanation to all terms and the method adopted:**

To know time from the Gnomonic shadow there are to terms involved for the calculation.

*(i)* *Madhya prabhā*
*(ii)* *Dinārdha danda*

Again for the calculation *of Madya prabhā* and *dinārdha danda* we need to calculate *Carārdha*, *Nadi* and *Nata*. *Nata* has two parts, *saumya nata* and *yamya nata*.

The first step of this method is to decide wheather the Sun is in northern or southern hemisphere. If Sun is in northern hemisphere then *akhya* has to be subtracted and will be added otherwise. It is because the author of *Bhasvati acharya Satananda* had made all calculations with reference to Puri, Odisha in northern hemisphere. Therefore when the Sun travels from northern to southern hemisphere It has to pass the equator the zero equinoxial gnomonic shadow line. Hence to consider the gnomonic shadow when Sun is in southern hemisphere a term *akhya* has to be added. According to Bhasvati Sun lies in northern hemisphere i.e. the days elapsed from vernal equinox to autumnal equinox is 187 days(modern data 186 days) and from autumnal equinox to vernal equinox is 178 days(modern data 179 days).

In second step we have to calculate *Carārdha*(spreading) . As we know day and night duration changes every day and it is not completely uniform. Therefore to take care of the changes in a day duration *Carārdha* has to be calculated. This method is an imperical method and Acharya Satananda claims that the method is completely of his own and he had not followed the advice from any previous scriptures. From madhy *prabhā* the maidday gnomonic shadow for the day concerned can be derived. From the proportion of *Madya prabhā* and *Iṣṭa chāyā* the time can be calculated

*Dinārdha danda* can be calculated by adding or subtracting *Carārdha* lita from the *dinārdha danda* on *Mahāviṣuva saṅkrānti* i.e. 15 danda dependening upon when Sun is in northern or southern hemisphere respectively. Comparision table for Midday gnomonic shadow on all 12 sangkrantis are with modern data has been given below .

The length of the shadow of the gnomon should be recorded of the moment of which the time has to be calculated. This is known as *iṣṭa chāyā*.

*iṣṭa chāyā* X 10 + 100 - *Madhya prabhā* = *Śaṅku* -----------------------*(1)*

(This *Śaṅku* is different from the gnomon itself)

Keep *dinārdha* (half day duration) of that day. Convert *daṇḍa* and *litā* into *litā* by multiplying 60 with danda and then adding *litā*. Now multiply *litā* pind with 100 and then divide it with the value of *Śaṅku* in equation (1). The result is the *iṣṭa chāyā kāla* (Time).

This time is of two types, *Gata kāla*: from morning upto noon and *Eṣva kāla*: from noon up to evening.

***Madhya prabhā* :** To know *Madhya prabhā* the *carārdha litā* is necessary to be calculated .

Then multiply 6 with *carārdha litā*. Keep the result in two places. Subtract one tenth of it from the number in second places. If the Sun is in northern hemisphere then keep the number as it is, else add one third of the number with it. Again divide the number with 10. If the Sun is in southern hemisphere then *akhya* has to be added.

***Carārdha***: Śatānanda claimed in his scripture that this method of calculation of *Carārdha* is completely of his own.

According to him if Sun is in Aries (*Meṣa*) , then the day count + the half of the day count is the *carārdha litā*. If Sun is in Tarus (*Vṛiṣa*) then *Carārdha* will be the *carārdha litā* of *Meṣa* + number of days elapsed from *Vṛiṣa* + one sixth of number of days elapsed from *Vṛiṣa*.

Again if Sun is on Gemini (*Mithuna*), the half of the days elapsed from the month of Mithuna has to be added with the *carārdha* of the month *Vṛiṣa*. The result is the *carārdha litā* for the month of Gemini (Mithuna). The *carārdha litā* for the month of *Karkaṭa* to *Kanyā* will decrease in the similar manner and on *Kanyā Saṅkrānti* it will be zero. Similar calculation has to be followed if the Sun is in southern hemisphere.

***Dinārdha* (Half day duration)**: The half day duration on *Mahāviṣuva saṅkrānti* is 15 *daṇḍa*. Calculate the *carārdha litā* for the day concerned. add the *carārdha litā* with 15 if Sun is in northern hemisphere and subtract if Sun is in southern hemisphere. The result is the required dinardha (half day duration) for the day concerned.

**Table-3 Midday Gnomonic shadow on all 12 Sangkranti**

| Sl No. | Declination of Sun (δ) in degrees | Right ascension of Sun (λ) in degrees | Midday gnomonic shadow from modern method | Midday gnomonic shadow from method in Bhasvāti | Difference= Error in % |
|---|---|---|---|---|---|
| 1. | 0.0 | 0.0 | 4.3676 | 4.45 | 0.0824=0.69% |
| 2. | 11.5008 | 30.0 | 1.7933 | 1.9788 | 0.1855=1.55% |
| 3. | 20.2017 | 60.0 | 0.04225 | 0.098 | 0.0557=0.46% |
| 4. | 23.5 | 90.0 | -0.7339 | -0.658 | 0.0759=0.63% |
| 5. | 20.2017 | 120.0 | -0.04225 | -0.037 | 0.0795=0.66% |
| 6. | 11.5003 | 150.0 | 1.7933 | 1.739 | -0.543=0.45% |
| 7. | 0.0 | 180.0 | 4.3676 | 4.45 | 0.082=0.69% |
| 8. | -11.5004 | 210.0 | 7.3537 | 7.496 | 0.1423=1.19% |
| 9. | -20.2017 | 240.0 | 10.1414 | 10.232 | 0.0906=0.75% |
| 10. | -23.5 | 270.0 | 11.3875 | 11.24 | -0.1475=1.23% |
| 11. | -20.2017 | 300.0 | 10.1414 | 10.148 | 0.0066=0.05% |
| 12. | -11.5008 | 330.0 | 7.3537 | 7.595 | 0.2413=2.025% |

Since Satananda has made calculation with respect to ahargana, so to have all calculation in same frame of reference I have adopted the data provided by NASA for my calculation. The old data table by NASA is given below, where March 21 has been taken as Mahabisuba Sangkranti or Mesha Sangkranti. In Bhāsvati it is also mentioned that Sun lies in northern hemisphere for 187 days and 178 days in southern hemisphere, which is same as the NASA table.

## Conclusion:

In this paper, the contribution of Śatānanda to the world of mathematics and astronomy has been discussed. Some of the *Ślokas* from his scripture *Bhāsvatī* has been translated to explain his achievements. It was necessary to prepare an accurate almanac for the Hindu society, mostly for the benefit of Jagannātha temple at *Puruṣottama dhāma* Purī. For this purpose he applied the observational data of Varāhamihira and took 450 C.E. the year when the scripture *Pañcasiddhāntikā* of Varāhamihira was written, as zero *ayanāṁśa*' year. Śatānanda started *Śastrābda* from the year he dedicated *Bhāsvatī* to the Society. All calculations in *Bhāsvatī* were in *Śastrābda* and he had given rules to convert *Śāstrābda* to *Śakābda* and vice versa. *Śatānanda* has taken the latitude and longitude of Puri, Odisha as his reference point. May be it was easy for him to recheck his methods from the observation at his native place.

The most interesting thing found in *Bhāsvatī* is that Śatānanda could calculate the position and rate of motion of heavenly bodies quite accurately without using trigonometric functions. Though some ancient astronomer had rejected the methodology by saying the method to be an approximate method, it is interesting to see that an approximate method could conclude with an exact solution of predetermining the eclipse. Use of *Śatāṁśa* (Centesimal system) in the procedure and making a back transform is quite modern idea adopted by Śatānanda. A strong claim exists that the conversion of the sexagecimal system to the centesimal system is the first step that led mathematicians towards the introduction of decimal system in mathematical calculations[1]. It is necessary to study the physical and mathematical interpretation of all 128 *ślokas* in *Bhāsvatī.*

A detail study is in progress to establish the relation among the method in *Bhāsvatī* and modern European method to predetermine eclipse.

## Appendix A: Method of calculation of sidereal period of Moon

Step -1. Multiply 90 with ahargana and add Chandra Dhruba with it. Divide the result with 2457.

Step-2 . Multiply 100 with ahargana and add Kendra dhruba with it. Divide the result with 2756.

Step-3. Divide ahargana with 120 and add the reminder of step 1. The carardha of the respective month has to be subtracted from the result.(carardha for different months are given below.)

step-4. Divide ahargana with 50 and add the reminder of step-2 . Then divide the result with 100.

Step-5. From the quotient the corresponding Khanda and Anukhanda (khanda +1) has to get from khanda table given below. Subtract Khanda from Anukhanda, the result is Chandra bhoga. Reminder from step-4 has to be multiplied by Chandra bhoga. Divide the resule with 100. The result has to be added with khanda and the result of Step-3. The result is Chandra Sphuta.

In the similar manner Chandra sphuta for the next day(ahargan) has to be calculated. The positional difference of the day is called Chandra bhukti( moon's durinal motion). This motion is not uniform. Therefore for the sidereal calculation I have kept on increasing the ahargana until moon's comes to the same position (Chandra Sphuta)

**Table for Carardha has to be subtracted in different months**

| Name of Sidereal Month | Carardha | Name of Sidereal Month |
|---|---|---|
| Aries | 0 | Pisces |
| Taurus | 1 | Aquarius |
| Gemini | 2 | Capricorn |
| Cancer | 2 | Sagittarius |
| Leo | 1 | Scorpio |
| Virgo | 0 | Libra |

### Table: Chandra Khanda-difference (antara) – Bhukti bodhaka Chakra

| | | | | | | | | | |
|---|---|---|---|---|---|---|---|---|---|
| **0** | **1** | **2** | **3** | **4** | **5** | **6** | **7** | **8** | **Number** |
| **0** | **0** | **1** | **3** | **6** | **10** | **16** | **24** | **35** | **Khanda** |
| **0** | **1** | **2** | **3** | **4** | **6** | **8** | **11** | **11** | **difference** |
| **9** | **10** | **11** | **12** | **13** | **14** | **15** | **15** | **17** | **Number** |
| **46** | **60** | **75** | **91** | **108** | **126** | **143** | **159** | **175** | **Khanda** |
| **14** | **15** | **16** | **17** | **18** | **17** | **16** | **16** | **15** | **Difference** |
| 18 | 19 | 20 | 21 | 22 | 23 | 24 | 25 | 26 | Number |
| 190 | 202 | 213 | 222 | 230 | 235 | 239 | 241 | 242 | **Khanda** |
| 12 | 11 | 9 | 8 | 5 | 4 | 2 | 1 | 1 | **Difference** |

| | | |
|---|---|---|
| 27 | 28 | Number |
| 243 | 243 | Khanda |
| 0 | 0 | Difference |

## Appendix B: Method of calculation for midday gnomonic shadow in different Sangkrantis

NASA table for different Sangkranti

(1)Aries Mar.21-April 20 =31
(2) Taurus April 21 – May 21=31
(3) Gemini May 22- June 21 = 31
(4) Cancer June 22 – July 22 =31
(5) Leo July 23 – Aug 21 =30
(6) Virgo Aug 22 – Sept. 23 =33

SUN LIES IN NORTHERN HEMISPHERE FOR 187 DAYS

(7) Libra Sept 24 – Oct 21 =28
(8) Scorpio Oct 22 – Nov 22 =32
(9) Sagittarius Nov 23 – Dec 22 =30
(10) Capricorn Dec. 23 – Jan 20 =29
(11) Aquarius Jan 21 – Feb 19 =30
(12) Pisces Feb 20 – March 20 =29

SUN LIES IN SOUTHERN HEMISPHERE FOR 178 DAYS

According to *Bhāsvatī* the pala prabha (equinoxial midday gnomonic shadow) is 4|27 =4.45

This is little higher than that of modern data. i.e. 4.37 +0.08

1. On March 21, Sun lies on the equator. So we take Sun's position in $0^0$. Aries
   So the gnomonic shadow will be 4.45.

2. On April 21 . : Taurus = $30^0$ = ahargana = 31 = 30 +1

*carārdha litā* = 45+1+1/6 = 46.17

46.17 X 6 =277.02 -27.70 = 249.32 /10 =24.932

madhy *prabhā*= 44.72 -24.932 = 19.788

ista chaya = 19.788/10=1.9788

3. May 22, Gemini : $60^0$ =ahargana 62 = 30 + 30 +2

   *carārdha litā* = 45+35+1 =81

81X 6 = 486 – 486/10 = 437.4/10 = 43.74

madhy *prabhā*= 44.72 - 43.74 = 0.98

ista chaya = 0.98/10=0.098

4. June 22 , Cancer : $90^0$ = ahargana 93 = 30 + 30 +33

   *carārdha litā* = 45+35+33/2 = 96.5

96.5 X 6 = 579 -57.9 = 521.1/10 = 52.11

madhy *prabhā*= 44.72 – 52.11 = - 7.39

ista chaya = - 7.39/10 = 0.739

5. July 23, Leo : $120^0$ =ahargana 124

(In this case there is little change in procedure. It has been mentioned that sun lies 187 days in northern hemisphere and 178 days in southern hemisphere. So when ahargana exceeds half of the days in a hemisphere then we have to take the smaller part for *carārdha litā* calculation. i.e.

187 – 124 = 63.

So we have to calculate the *carārdha litā* of 63 ahargana.)

63 = 30 + 30 +3

*carārdha litā* = 45+35+3/2 = 81.5

81.5 X 6 = 501- 50.1= 450.9/10 = 45.09

madhy *prabhā*= 44.72 – 45.09 = -0.37

ista chaya = - 0.37/10= -0.037

6. Aug 22, Virgo : $150^0$ = ahargana 154= 187-154 = 33=30 + 3

*carārdha litā* = 45+3+3/6= 48.5

48.5 X 6 = 291-29.1 =261.9/10 = 26.19

madhy *prabhā*= 44.72 – 26.19 = 18.53

ista chaya = 18.53/10=1.853

7. Sept 24, Libra : $180^0$ = ahargana 187

Shadow length = 4.45

8. Oct 22, Scorpio : $210^0$ = ahargana 215

215-187 = (southern hemisphere)=28

*carārdha litā* = 28+14 = 42

(there is little change in procedure for southern hemisphere)

42 X 6 = 252-25.2 = 226.8 + 226.8/3 = 302.4/10=30.24

madhy *prabhā*= 44.72 + 30.24= 74.96

ista chaya = 74.96/10=7.496

9. Nov 23, Sagittarius : $240^0$ = ahargana 247

247-187 = 60

*carārdha litā* = 45 +35 =80

80 X 6 =480 – 48 =432 + 432/3 = 576/10 =57.6

madhy *prabhā*= 44.72 + 57.6 = 102.32

ista chaya = 102.32/10=10.232

10. Dec. 23, Capricorn: $270^0$ = ahargana 277

277-187 = 90

southern hemisphere 178 -90 = 88

We have to calculate *carārdha litā* of the smaller part.

So *carārdha litā of 88* =45 +35 +14 = 94

94 X 6 =564 – 56.4 = 507.6 + 507.6/3 = 676.8/10 = 67.68

madhy *prabhā*= 44.72 + 67.68 = 112.4

ista chaya = 112.4/10=11.24

11. Jan 21, Aquarius : $300^0$ = ahargana 306

306 – 187 = 119

178 – 119 = 59

59 = 45 + 29 +29/6 =78.83

78.83 X 6 = 473 – 47.3 = 425.7 + 425.7/3 = 567.6/10 = 56.76

madhy *prabhā*= 44.72 + 56.76 = 101.48

ista chaya = 101.48/10=10.148

12. Feb 20, Pisces :330 = ahargana 336

336 -187 = 149

178 – 149 = 29

29+ 29/2 = 43.5

43.5 X 6 = 261 – 26.1 = 234.9 + 234.9/3 = 312.3/10 = 31.23

madhy *prabhā* =44.72 + 31.23 = 75.95

Ista chaya =75.95/10 = ***7.595***